%% file: main.tex
 \documentclass[reprint,onecolumn,PRF]{revtex4-2} 

\usepackage[utf8]{inputenc}
\usepackage{amsmath}

\usepackage{graphicx}
\usepackage{amsmath,amssymb,amstext} 

\input{MacrosSymbols}

\begin{document}

\title{Accounting for surface temperature variations in Rayleigh-B\'{e}nard convection}
\author{Jason Olsthoorn}
\affiliation{Department of Civil Engineering, Queen's University, Kingston, Ontario, Canada, K7L 3N6 }
\email[]{Jason.Olsthoorn@queensu.ca}

\begin{abstract}
Turbulent Rayleigh-B\'{e}nard convection is often modelled with a constant surface temperature. However, the surface temperature of many geophysical systems, such as lakes, is coupled to the atmospheric forcing. In this paper, we account for this dynamic surface temperature through an additional parameter $\beta$. Using an appropriately defined dynamical Rayleigh number $\RaD$, we recover many of the results from the standard Rayleigh-B\'{e}nard model. We hope that this work will simplify the application of Rayleigh-B\'{e}nard theory in geophysical contexts, such as lakes.
\end{abstract}


\maketitle

\section{Introduction}

In its usual configuration, Rayleigh-B\'{e}nard convection results from sufficiently heating the bottom, and cooling the top, of a fluid. 
In their experimental work \citep{benard_les_1901}, B\'{e}nard observed that the surface water deflections from this convection formed hexagonal cells.
In an attempt to model these cells, \citet{lord_rayleigh_convection_1916} then made the simplification that ``the fluid is supposed to be bounded by two infinite fixed planes [...], where also the temperatures are maintained constant." Lord Rayleigh noted that this was a simplification from ``B\'{e}nard's experiments, where, indeed, the [temperature] conditions are different at the two boundaries." In the years since that publication, Rayleigh-B\'{e}nard convection has proven to be a complex dynamical system and remains a topic of intense academic interest in fields ranging from astrophysics to oceanography \citep{doering_turning_2020}. 

In many geophysical systems, such as lakes and oceans, the water surface can be heated or cooled from the surrounding environment. For example, the surface temperature of a lake depends on the atmospheric temperature, surface radiation, and evaporation. During Autumn, the lake surface cools, which drives convection that will transport heat from within the lake to the water surface. That is, the surface temperature depends on both the atmospheric forcing and the convection; it is not a constant as is typically considered in Rayleigh-B\'{e}nard convection. However, we will show that this coupling can be encapsulated by the Biot number, $\beta$, and that, with appropriately defined dynamic parameters, the results of this modified setup are similar to those found in the standard Rayleigh-B\'{e}nard model. Thus, we hope that this work will simplify the application of Rayleigh-B\'{e}nard theory in geophysical contexts, such as lakes. 

Modifying the surface boundary conditions will change the linear stability of the Rayleigh-B\'{e}nard model, which has been previously discussed in \citet{sparrow_thermal_1964} and subsequently in \citet{foster_effect_1968}. The linear stability of the system is determined by the Rayleigh number ($\Ra$), which characterizes the ratio of advective to diffusive transport. \citet{sparrow_thermal_1964} demonstrated that the critical Rayleigh number (Ra$_C$), the minimum Rayleigh number for instability, monotonically increases as the upper boundary condition changed from an insulating condition ($\beta=0$, Ra$_C\approx720$) to an isothermal condition ($\beta\to\infty$, , Ra$_C\approx1800$). Similarly changing the upper velocity boundary condition from free-slip to no-slip resulted in a near uniform decrease in Ra$_C$ ( a decrease of $\approx100$, see figure 1a in \citet{sparrow_thermal_1964}). However, while the boundary conditions affect Ra$_C$, \citet{foster_effect_1968} demonstrated that the precise value of the thermal boundary condition has a weak  effect on the time to instability ($t_0$) for the semi-infinite system. In this paper, we will focus on the nonlinear behaviour of the system, after it becomes convectively unstable, both before and at thermal equilibrium.  

Once started, convection enhances the effective thermal conductivity between the two boundaries in the Rayleigh-B\'{e}nard system, which increases with $\Ra$. 
As a result, \citet{chilla_ultimate_2004} argue that, at high $\Ra$, the finite thermal conductivity of the bounding plates used in laboratory experiments will be unable to maintain a fixed temperature. That is, the true boundary temperature in laboratory experiments depends upon the convection once the effective thermal conductivity of the fluid increases beyond a certain value. 
Subsequently, \citet{verzicco_effects_2004} and \citet{brown_heat_2005} demonstrated that an empirical correction factor that depends on the conductivity ratio between the plates and the fluid can account for the relative decrease in laboratory-measured heat transport at high $\Ra$. Interestingly, \citet{wittenberg_bounds_2010} argue that where the finite conductivity of the plates is significant (as is the case when $\Ra\to\infty$), the appropriate $\Ra$ is defined by the temperature difference across the entire system, including the conductive plates. We will define $\Ra$ in a similar manner below. While the methods and application of this paper are different from these laboratory setups, we will similarly argue that the ratio $\beta/\Ra^p$, ($p$ is a constant defined below) determines if the surface temperature is effectively ``fixed" at a constant temperature, or if the convection significantly modifies the surface temperature.

More recently, \citet{clarte_effects_2021} performed a set of three-dimensional convective simulations of a semi-insulated rotating convective shell with a range of Biot numbers $\beta$. They argue that the dynamic surface boundary condition can be replaced by a fixed flux condition when $\beta$ is sufficiently small, or replaced by a fixed temperature condition when $\beta$ is sufficiently large. In this paper, we will extend their work by arguing that the ratio $\beta/\Ra^p$ is the important parameter that determines the surface boundary value.

We present a model of convection between two fixed planes with a dynamic surface boundary condition for temperature. This boundary condition will assume that the far-field atmospheric conditions are fixed, with a dynamic thermal boundary-layer near the water surface. We want to answer the following key questions:
\begin{enumerate}
\setlength{\itemsep}{0pt}
  \setlength{\parskip}{0pt}
  \setlength{\parsep}{0pt}
    \item What is the heat transfer rate at the water surface?
    \item What is the equilibrium surface water temperature? 
    \item How quickly does the system reach equilibrium?
    \item How vigorous is the convection? 
\end{enumerate}
In the following discussion, we will describe the problem setup (\S \ref{sec:ProblemSetup}), and the theoretical framework to answer the questions above (\S \ref{sec:Theory}). The results of the 2D numerical simulations will be discussed in \S \ref{sec:Results}, before concluding in \S\ref{sec:Conclusions}.

\section{Problem Setup}\label{sec:ProblemSetup}

In analogy to a surface cooled lake, we model a body of water that is cooled by exposure to the atmosphere through the top boundary. In this study, we will fix the bottom temperature of the water to a known value $\TzB$. Further, we consider the case where the water is shallow compared to its width, which we will model as periodic. This is a very similar setup to the classic Rayleigh-B\'{e}nard problem with one major modification: the surface temperature is coupled to the flux of heat through the surface. 

\begin{figure}
    \centering
    \includegraphics[width=0.95\textwidth]{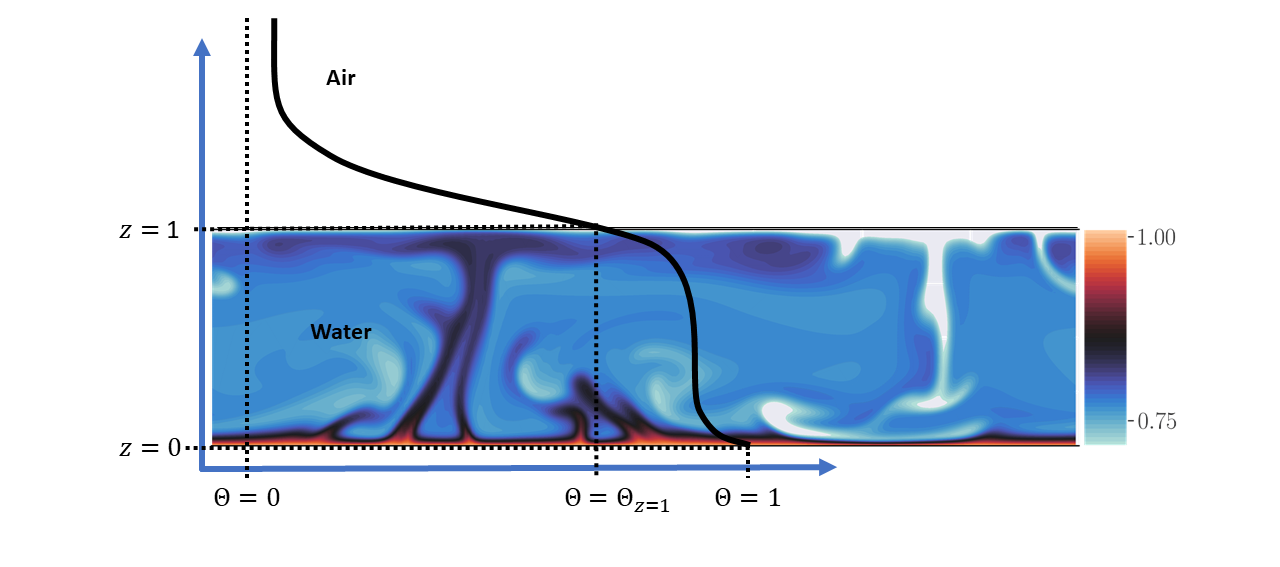}
    \caption{A contour plot of the simulated value of nondimensional temperature $\Theta$ for $\Ra=10^7,\beta=8$ at time t=1500. Superimposed is a cartoon of the mean temperature profile ($\Havg{\Theta}$) between the fixed tank bottom at $\Theta=1$ and the air temperature above.  }
    \label{fig:diagram}
\end{figure}

Figure \ref{fig:diagram} is a schematic of the non-dimensionalized (see below) mean temperature profile $\left(\mTh\right)$ in this configuration. The temperature is fixed at the bottom of the domain, which is warmer than the atmosphere above. Due to the resultant convection, the temperature at the centre of the water domain is well mixed, with temperature boundary layers at the top and bottom boundary. The surface water temperature depends on how much heat is being transported upwards by the convection.

\subsection{Temperature Boundary Condition}

The rate of change of the dimensional mean water temperature $\tilde T\ [\hbox{K}]$ in a volume $V\ [\hbox{m}^3]$ is determined by the heat flux $\mathbf{q}\ [\hbox{W m}^{-2}]$ through the boundaries $A\ [\hbox{m}^2]$. That is, 
\begin{align}
   \deriv{\tilde t}{} \intV{\rho_0 c_p \tilde T} &= \intA[A]{-\mathbf{q}\cdot \hat {\mathbf{n}}}. \label{eqn:heatBudget}
\end{align}
Here, $\hat{\mathbf{n}}$ is the outward normal vector, $\rho_0\ [\hbox{kg m}^{-3}]$ is the reference density of water, and $c_p\ [\hbox{J kg}^{-1} \hbox{K}^{-1}]$ is the specific heat capacity of water. As mentioned above, we fix the lower boundary temperature ($\TzB$). The domain is horizontally periodic. The heat flux at the surface depends on the driving environmental process.

In many natural systems, such as lakes and oceans, the surface heat loss is given by the sum of sensible (conductive), radiative, and latent heat fluxes. In this paper, we omit any additional heat inputs, such as precipitation, and we will ignore short-wave radiation as it is not strictly a boundary effect. The surface heat flux is then computed as the sum of net longwave radiation, sensible heat flux, and evaporation, written:
\begin{gather}
    \Qs\cdot \mathbf{n} = -\epsilon_w \sigma_{SB} \TzH^4  + \epsilon \sigma_{SB} \TA^4 - K_{Air} \left( \TzH - \TA\right) + q_L.
\end{gather}
Here, $\sigma_{SB}\ [\hbox{W m}^{-2} \hbox{K}^{-4}]$ is the Stefan-Boltzmann constant and $\TA$ is the atmospheric temperature. The effective air heat transfer coefficient $\left(K_{Air}\ [\hbox{W m}^{-2} \hbox{K}^{-1}]\right)$, and the emissivity of the air ($\epsilon$ [-], including reflection from the water surface) and water $\left(\epsilon_w\ [-] \right) $ are model parameters and need to be computed. However, these parameters have been studied extensively in the literature (see \citet{lerman_mixing_1995} and elsewhere). Finally, the heat loss due to evaporation is denoted as $q_L\ [\hbox{W m}^{-2}]$. 

\citet{hitchen_impact_2016} showed that in the absence of evaporation ($q_L=0$), the surface heat flux ($\Qs$) can be linearized to the following:
\begin{gather}
    \Qs\cdot \mathbf{n} =  \gamma \left( \TzH - \TO\right), 
    \label{eqn:Qs}
\end{gather}
where, in our notation,  
\begin{align}
    \gamma =   \left( K_{Air}    + 4 \epsilon_w \sigma_{SB}  \TA^3 \right), \qquad \TO = \TA + \frac{\sigma_{SB} \left(\epsilon - \epsilon_w \right)\TA^4}{\left( K_{Air}    + 4 \epsilon_w \sigma_{SB}  \TA^3 \right)}. 
\end{align}
Here, $\gamma\ [\hbox{W m}^{-2} \hbox{K}^{-1}]$ is an effective heat transfer coefficient and $\TO$ is an effective reference temperature.
This approximation assumes that the surface water temperature $\TzH$ and the atmospheric temperature $\TA$ are close $(|\TzH - \TA | \ll \TA$, in Kelvin). 
This model can be similarly written to include the effects of evaporation, which modify $\gamma$ and $\TO$, but not the functional form of \eqref{eqn:Qs}. We do not include them here as they increase the complexity of the equations, but we refer the interested reader to \citet{lepparanta_freezing_2015}. For the purposes of this paper, we will assume that $\gamma$ and $\TO$ are given.

\subsection{Equations of motion}

The difference between the reference temperature $\TO$ and the bottom temperature $\TzB$ can be used to define a characteristic velocity $U_0^2={g H \alpha \left(\TzB - \TO\right)}$, where $\alpha$ is the assumed constant thermal expansion coefficient, $g$ is the acceleration due to gravity, and $H$ is the water depth. We scale length by $H$, velocity ($\mathbf{u}=(u,w)$) by $U_0$, pressure by $\rho_0 U_0^2$, and time by the advective timescale $\tau_0 = H/U_0$. Temperature is similarly scaled as
\begin{gather}
    \Theta = \frac{\tilde T - \TO}{ \TzB - \TO}. 
\end{gather}
Under this non-dimensionalization, the equations of motion are
\begin{gather}
    \left( \pderiv{t}{} + \mathbf{u} \cdot \nabla \right)\mathbf{u} = - \nabla P  +  \Theta \mathbf{\hat k} + \sqrt{\frac{\Pr}{\Ra}} \nabla^2 \mathbf{u},
	\label{eqn::momentum}\\
	\left( \pderiv{t}{} + \mathbf{u} \cdot \nabla \right) \Theta = \frac{1}{\sqrt{\Ra\ \Pr}}\nabla^2 \Theta, \label{eqn:Temp} \\ 
	\nabla \cdot \mathbf{u} = 0. \label{eqn::DivFree}
\end{gather}
These equations contain two non-dimensional parameters: the Rayleigh number ($\Ra$) and the Prandtl number (Pr). 
\begin{gather}
    \Ra =  g \alpha \left( \TzB - \TO\right)\frac{H^3}{\kappa \nu}, \qquad \hbox{Pr} = \frac{\nu}{\kappa}.
\end{gather}
Here, $\nu$ is the kinematic viscosity and $\kappa$ is the thermal diffusivity.

We model the top and bottom boundaries as impermeable that are free-slip (no tangential stress) at the surface and no-slip at the bottom. 
The boundary conditions are then written
\begin{align}
    \pderiv{z}{u} = 0, \quad& w = 0, & \pderiv{z}{\Theta} &= -\beta \Theta, \qquad \hbox{ where }\beta = \frac{\gamma H}{ \rho_0 c_p \kappa}, & z=1, \label{eqn:BC}\\
    u=0, \quad& w = 0,& \Theta &= 1, & z=0.
\end{align}
  For simplicity, the initial condition was set to $\Theta(z) = 1$. Note that the numerical solver instantaneously corrects the initial condition to match the boundary condition. 
  We perform a series of numerical simulations with different values of $\Ra$ and $\beta$. In all cases, we fix $\Pr=9$ as the approximate value for heat in water ($\Pr\approx9.0$ at 11.5$^\circ$C and atmospheric pressure at sea level).

\subsection{Numerical Methods}
We solve the system of equations \eqref{eqn::momentum}--\eqref{eqn::DivFree} with Dedalus \citep{burns_dedalus_2020}, using pseudospectral spatial derivatives (Chebyshev polynomials in the vertical and Fourier modes in the horizontal) and a second-order Runge-Kutta time-stepping scheme. Numerical convergence was verified with grid resolution studies and by ensuring that there were at least 8 grid points within the top boundary layer.  

In total, 44 numerical simulations were run (see table~\ref{tab:CaseParameters}). The simulations were two-dimensional and run for a sufficiently long time that the system reached a quasi-steady-state (final time was selected with $t/\tau\gtrsim4$ as defined below). The domain width was 4 times the depth in all cases.

\begin{table}
	\begin{center}
		\begin{tabular}{c c |c c c c |c c c c |c c c c c}
$\Ra$ & $\beta$ & Nx & Nz & $\delta_{top}$ & $N_B$ & $t_0$ & $\epsilon$ & $t_{max}$ & $t_{max}/\tau$ & $\quad$ Nu$_0$ $\quad$ & $\Theta_1$ & $\sigma(\Theta_1)$ & $\TKE_0$ & Re \\
			\hline
			$10^{6}$ & $2^{0}$ & 256 & 128 & 0.034 & 15 & 15.7 & 0.058 & 1700 & 4.8 & 6.3 & 0.9 & 0.02 & 1.1$\times 10^{-3}$ & 15 \\
			$10^{6}$ & $2^{1}$ & 256 & 128 & 0.030 & 14 & 11.7 & 0.115 & 1300 & 4.4 & 7.4 & 0.8 & 0.03 & 1.7$\times 10^{-3}$ & 19 \\
			$10^{6}$ & $2^{2}$ & 256 & 128 & 0.030 & 14 & 9.3 & 0.230 & 1100 & 4.4 & 8.4 & 0.7 & 0.04 & 2.8$\times 10^{-3}$ & 24 \\
			$10^{6}$ & $2^{3}$ & 256 & 128 & 0.028 & 14 & 7.1 & 0.460 & 1300 & 6.2 & 8.9 & 0.5 & 0.05 & 4.3$\times 10^{-3}$ & 30 \\
			$10^{6}$ & $2^{4}$ & 256 & 128 & 0.025 & 13 & 6.5 & 0.920 & 1200 & 6.8 & 9.4 & 0.4 & 0.05 & 6.1$\times 10^{-3}$ & 36 \\
			$10^{6}$ & $2^{5}$ & 256 & 128 & 0.026 & 13 & 4.5 & 1.841 & 1100 & 7.4 & 10.3 & 0.2 & 0.05 & 7.8$\times 10^{-3}$ & 41 \\
			$10^{6}$ & $2^{6}$ & 256 & 128 & 0.026 & 13 & 3.7 & 3.681 & 1100 & 8.8 & 10.2 & 0.1 & 0.04 & 9.1$\times 10^{-3}$ & 44 \\
			$10^{6}$ & $2^{7}$ & 256 & 128 & 0.024 & 13 & 3.6 & 7.362 & 1000 & 9.7 & 10.3 & 0.1 & 0.03 & 9.7$\times 10^{-3}$ & 46 \\
			$10^{6}$ & $2^{8}$ & 256 & 128 & 0.026 & 13 & 3.4 & 14.724 & 1000 & 9.7 & 10.3 & 0.0 & 0.02 & 10.0$\times 10^{-3}$ & 47 \\
			$10^{6}$ & $2^{9}$ & 256 & 128 & 0.026 & 13 & 3.4 & 29.448 & 1000 & 9.7 & 10.3 & 0.0 & 0.01 & 1.0$\times 10^{-2}$ & 48 \\
			$10^{6}$ & Fixed & 256 & 128 & 0.025 & 13 & 2.7 & $\infty$ & 1000 & 9.7 & 10.5 & 0.0 & 0.00 & 1.1$\times 10^{-2}$ & 48 \\
			\hline 
			$10^{7}$ & $2^{0}$ & 256 & 128 & 0.025 & 13 & 18.4 & 0.028 & 3600 & 5.5 & 10.4 & 0.9 & 0.01 & 9.4$\times 10^{-4}$ & 45 \\
			$10^{7}$ & $2^{1}$ & 256 & 128 & 0.022 & 12 & 15.0 & 0.056 & 3000 & 5.5 & 11.9 & 0.9 & 0.02 & 1.7$\times 10^{-3}$ & 61 \\
			$10^{7}$ & $2^{2}$ & 256 & 128 & 0.019 & 11 & 10.8 & 0.113 & 2400 & 5.2 & 13.6 & 0.8 & 0.03 & 3.0$\times 10^{-3}$ & 81 \\
			$10^{7}$ & $2^{3}$ & 256 & 128 & 0.016 & 10 & 7.3 & 0.225 & 2000 & 5.2 & 15.3 & 0.7 & 0.04 & 4.8$\times 10^{-3}$ & 103 \\
			$10^{7}$ & $2^{4}$ & 256 & 128 & 0.016 & 10 & 6.2 & 0.451 & 1600 & 4.9 & 16.8 & 0.5 & 0.05 & 7.8$\times 10^{-3}$ & 131 \\
			$10^{7}$ & $2^{5}$ & 256 & 128 & 0.016 & 10 & 4.7 & 0.901 & 1600 & 5.8 & 17.8 & 0.4 & 0.06 & 1.1$\times 10^{-2}$ & 156 \\
			$10^{7}$ & $2^{6}$ & 512 & 128 & 0.014 & 10 & 3.7 & 1.803 & 1500 & 6.5 & 18.5 & 0.2 & 0.06 & 1.3$\times 10^{-2}$ & 173 \\
			$10^{7}$ & $2^{7}$ & 512 & 128 & 0.014 & 10 & 3.2 & 3.606 & 1450 & 7.5 & 19.0 & 0.1 & 0.04 & 1.6$\times 10^{-2}$ & 186 \\
			$10^{7}$ & $2^{8}$ & 512 & 128 & 0.013 & 9 & 2.9 & 7.212 & 1400 & 8.7 & 19.3 & 0.1 & 0.03 & 1.7$\times 10^{-2}$ & 193 \\
			$10^{7}$ & $2^{9}$ & 512 & 128 & 0.014 & 9 & 2.7 & 14.423 & 1450 & 9.0 & 19.2 & 0.0 & 0.02 & 1.7$\times 10^{-2}$ & 196 \\
			$10^{7}$ & Fixed & 256 & 128 & 0.013 & 9 & 2.4 & $\infty$ & 1450 & 9.0 & 19.2 & 0.0 & 0.00 & 1.8$\times 10^{-2}$ & 199 \\
			\hline 
			$10^{8}$ & $2^{0}$ & 256 & 128 & 0.016 & 10 & 22.9 & 0.014 & 7000 & 5.8 & 17.3 & 0.9 & 0.01 & 9.1$\times 10^{-4}$ & 142 \\
			$10^{8}$ & $2^{1}$ & 256 & 128 & 0.013 & 9 & 15.9 & 0.028 & 5400 & 5.3 & 19.9 & 0.9 & 0.01 & 1.7$\times 10^{-3}$ & 195 \\
			$10^{8}$ & $2^{2}$ & 512 & 128 & 0.009 & 8 & 11.9 & 0.055 & 4800 & 5.6 & 22.4 & 0.8 & 0.02 & 2.9$\times 10^{-3}$ & 254 \\
			$10^{8}$ & $2^{3}$ & 512 & 256 & 0.006 & 13 & 7.9 & 0.110 & 4100 & 5.7 & 28.6 & 0.8 & 0.03 & 4.9$\times 10^{-3}$ & 329 \\
			$10^{8}$ & $2^{4}$ & 512 & 256 & 0.006 & 13 & 6.9 & 0.221 & 3300 & 5.5 & 32.2 & 0.7 & 0.04 & 8.1$\times 10^{-3}$ & 423 \\
			$10^{8}$ & $2^{5}$ & 512 & 256 & 0.006 & 12 & 4.9 & 0.442 & 2700 & 5.3 & 35.7 & 0.5 & 0.06 & 1.2$\times 10^{-2}$ & 521 \\
			$10^{8}$ & $2^{6}$ & 512 & 256 & 0.006 & 12 & 3.9 & 0.883 & 2300 & 5.4 & 38.7 & 0.4 & 0.06 & 1.7$\times 10^{-2}$ & 610 \\
			$10^{8}$ & $2^{7}$ & 512 & 256 & 0.006 & 12 & 2.9 & 1.766 & 2300 & 6.4 & 40.1 & 0.2 & 0.06 & 2.1$\times 10^{-2}$ & 682 \\
			$10^{8}$ & $2^{8}$ & 512 & 256 & 0.005 & 12 & 2.9 & 3.532 & 2100 & 7.0 & 41.5 & 0.1 & 0.05 & 2.4$\times 10^{-2}$ & 736 \\
			$10^{8}$ & $2^{9}$ & 512 & 256 & 0.006 & 12 & 2.9 & 7.064 & 2000 & 8.1 & 42.1 & 0.1 & 0.03 & 2.6$\times 10^{-2}$ & 765 \\
			$10^{8}$ & Fixed & 512 & 256 & 0.006 & 13 & 1.9 & $\infty$ & 2000 & 8.1 & 42.4 & 0.0 & 0.00 & 2.8$\times 10^{-2}$ & 795 \\
			\hline 
			$10^{9}$ & $2^{0}$ & 512 & 256 & 0.006 & 13 & 23.9 & 0.007 & 14500 & 6.5 & 27.2 & 1.0 & 0.00 & 7.3$\times 10^{-4}$ & 403 \\
			$10^{9}$ & $2^{1}$ & 512 & 256 & 0.005 & 11 & 16.9 & 0.014 & 11900 & 6.3 & 36.5 & 0.9 & 0.01 & 1.4$\times 10^{-3}$ & 550 \\
			$10^{9}$ & $2^{2}$ & 512 & 256 & 0.004 & 11 & 12.9 & 0.027 & 9300 & 5.9 & 42.4 & 0.9 & 0.01 & 2.5$\times 10^{-3}$ & 744 \\
			$10^{9}$ & $2^{3}$ & 1024 & 256 & 0.004 & 11 & 9.9 & 0.054 & 7000 & 5.3 & 48.4 & 0.9 & 0.02 & 4.3$\times 10^{-3}$ & 980 \\
			$10^{9}$ & $2^{4}$ & 1024 & 256 & 0.004 & 10 & 6.9 & 0.108 & 6300 & 5.6 & 54.2 & 0.8 & 0.03 & 7.3$\times 10^{-3}$ & 1273 \\
			$10^{9}$ & $2^{5}$ & 1024 & 256 & 0.004 & 10 & 4.9 & 0.216 & 5000 & 5.3 & 60.0 & 0.7 & 0.04 & 1.1$\times 10^{-2}$ & 1579 \\
			$10^{9}$ & $2^{6}$ & 1024 & 256 & 0.004 & 10 & 3.9 & 0.432 & 3900 & 4.9 & 65.9 & 0.5 & 0.05 & 1.6$\times 10^{-2}$ & 1897 \\
			$10^{9}$ & $2^{7}$ & 1024 & 256 & 0.004 & 10 & 2.9 & 0.865 & 3000 & 4.5 & 71.0 & 0.4 & 0.05 & 2.2$\times 10^{-2}$ & 2189 \\
			$10^{9}$ & $2^{8}$ & 1024 & 256 & 0.004 & 10 & 2.9 & 1.730 & 2400 & 4.3 & 74.9 & 0.2 & 0.05 & 2.7$\times 10^{-2}$ & 2441 \\
			$10^{9}$ & $2^{9}$ & 1024 & 256 & 0.004 & 10 & 1.9 & 3.460 & 1900 & 4.0 & 77.6 & 0.1 & 0.04 & 3.1$\times 10^{-2}$ & 2629 \\
			$10^{9}$ & Fixed & 1024 & 256 & 0.004 & 10 & 1.3 & $\infty$ & 1900 & 4.9 & 76.4 & 0.0 & 0.00 & 3.8$\times 10^{-2}$ & 2915 \\
			\hline
		\end{tabular}
	\end{center}
	\caption{A table of the parameters associated with each experiment. The horizontal and vertical grid resolution was $Nx,Nz$, respectively. The upper thermal boundary layer thickness was $\delta_{top}$, with $N_B$ grid points resolving the boundary layer. The initial time of instability is $t_0$ as defined as the time of minimum kinetic energy $\TKE$. The end time of each numerical run was $t_{max}$, which is at least $\ge4\tau$, where $\tau = \tau_\epsilon$ is used for $\epsilon<5$ and $\tau=\tau_\infty$ otherwise (both defined below). The estimated steady state values of $\Nu_0$, $\ThO$ and $\TKE_0$ are provided. The standard deviation of the surface temperature $\sigma(\ThO)$ is also provided. Finally, the Reynolds number (Re) based upon the RMS velocity is given. The `Fixed' cases have prescribed $\Tzt=0$, i.e. the limit of $\beta\to\infty$. The horizontal dimension Lx$ = 4$ in all cases. }
	\label{tab:CaseParameters}
\end{table}

\section{Theory}\label{sec:Theory}

The Nusselt number ($\Nu$) is the ratio between the measured surface vertical heat flux ($J$) and the purely diffusive heat flux. In our non-dimensionalization, it is computed as
\begin{gather}
    \Nu = \frac{ J}{ \Delta },
\end{gather}
 The temperature difference, $\Delta =  \Tzb - \Tzt $, between the upper and lower boundaries is typically prescribed for classical Rayleigh-B\'{e}nard convection.
For a fixed Prandtl number, we will show below that $\Nu$ scales with the effective Rayleigh number $\RaD$,
\begin{gather}
    \Nu - 1 \sim C \RaD^p, \qquad \RaD = \Ra \Delta, \label{eqn:NuScl}
\end{gather}
where the effective Rayleigh number, $\RaD$, is defined by the dynamic temperature values at the top and bottom boundary. 
While debate persists concerning the `ultimate' limit of the value of $p$ \citep{plumley_scaling_2019}, it is often shown to have a value of $p\approx 0.3$ at moderate Rayleigh numbers 
(e.g. 
\citet{niemela_turbulent_2006} estimate that $p\approx\frac{1}{3}$ for $\hbox{Ra}\approx10^{10}-10^{12}$,
\citet{plumley_scaling_2019} estimate that $p\approx0.322$ for $\hbox{Ra}\approx10^{5}-10^{15}$, 
\citet{chilla_ultimate_2004} estimate $p\approx0.3$ for $\hbox{Ra}\approx10^{9}-10^{12}$ ).

In steady state, the horizontally averaged ($\Havg{\left(\cdot\right)} = \frac{1}{A} \intA{\left(\cdot\right)}$) temperature equation~\eqref{eqn:Temp} simplifies to
\begin{gather}
    \diffz{} \left( \frac{1}{\sqrt{\Ra \Pr}} \diffz{\mTh} - \Havg{w\Theta} \right)=0, 
\end{gather}
where $w$ is the vertical velocity. Integrating the above equation, we show that the Nusselt number can be equivalently determined by the surface heat flux, or the volume averaged vertical heat flux as 
\begin{gather}
    \Nu  = \frac{1}{\Delta} \diffz{\mTh}\bigg|_{z=1} = \frac{\sqrt{\Ra\ \Pr}}{\Delta} \Vavg{w\Theta} + 1, \label{eqn:Nu}
\end{gather}
where $\Vavg{\left(\cdot\right)} = \frac{1}{V} \intV{\left(\cdot\right)}$). In the results presented below, we will numerically evaluate the Nusselt number based upon the surface gradient. We will use these relationships below. 

\subsection{Diffusive solution}
The simplest solution to equations \eqref{eqn::momentum}-\eqref{eqn::DivFree} with boundary conditions \eqref{eqn:BC} is the diffusive (non-convecting) solution. That is, if we make the ansatz that 
\begin{gather}
    \Theta = a z + 1 \implies a = - \beta \left(a + 1\right),\\
    \Theta = -\frac{\beta}{ \beta + 1} z + 1. 
\end{gather}
Therefore, for the purely diffusive problem, 
\begin{align}
    \Tzt = -\frac{\beta }{\beta  + 1}  + 1 
     \approx  \frac{1}{\beta } + \dots, \qquad \beta\to\infty \label{eqn:DiffSolution}
\end{align}
The surface temperature $\Tzt$ decreases with increasing $\beta$. For large $\Ra$, convection increases the vertical heat flux, resulting in a nonlinear temperature profile. Nonetheless, this diffusive solution highlights the approximate functional dependence of the surface temperature on $\beta$. 

\section{Results}\label{sec:Results}

\begin{figure}
        \centering
        \includegraphics[width=0.95\textwidth]{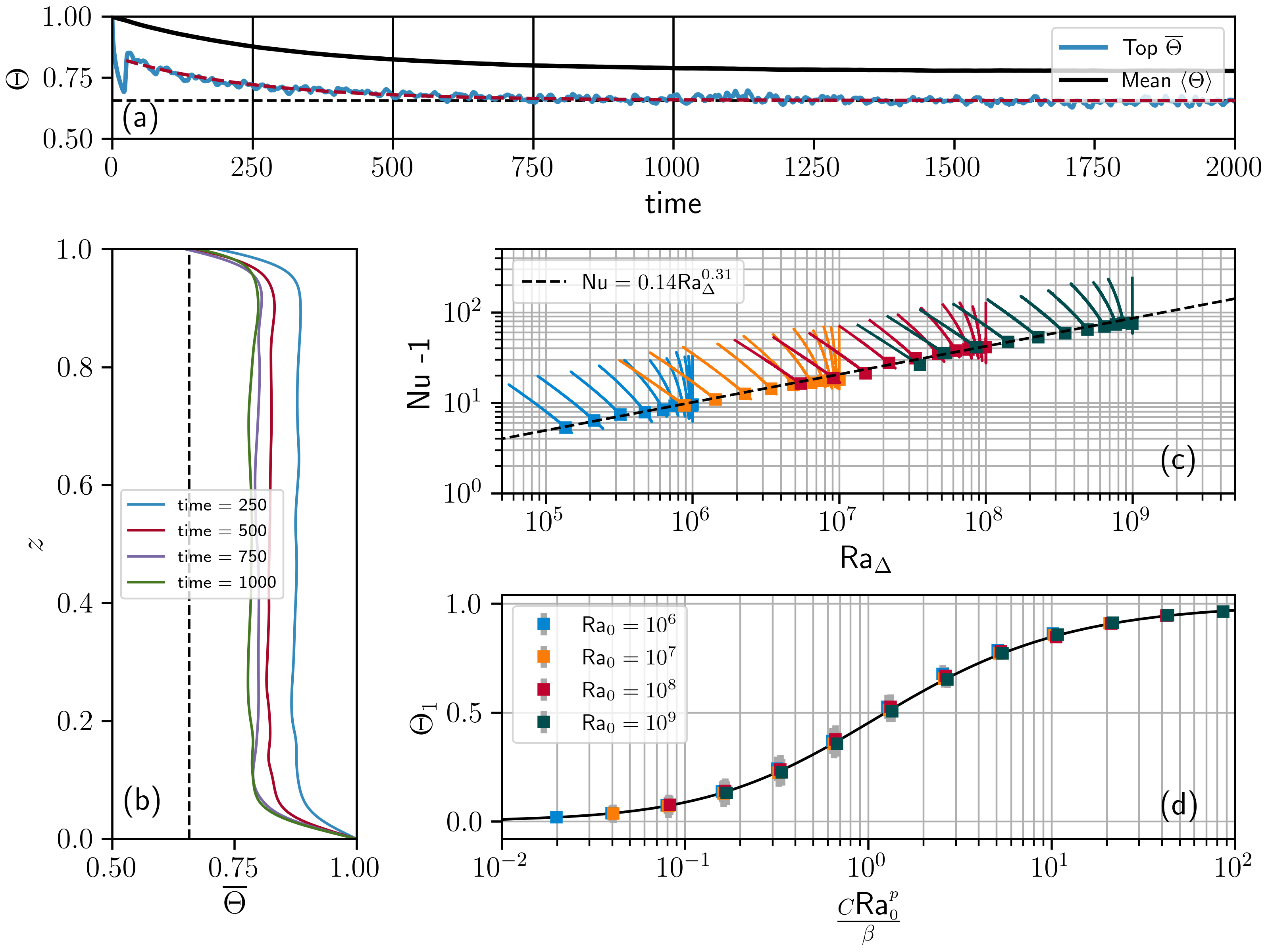}
        \caption{(a) The time series of the mean top and mean domain temperatures. The red dashed line is an exponential fit to the data. Note that these temperatures asymptote to statistically stationary values. (b) Mean temperature profiles at the times identified by vertical lines in (a). The data in (a) and (b) are from $\Ra = 10^7,\ \beta = 2^3$. (c) The Nusselt number as a function of $\RaD$. The lines denote the time evolution of $\Nu-1$ over the convective period of the simulation, while the squares denote the equilibrium values. (d) The asymptotic values of $\ThO$ as a function of $\RaD$ and $\beta$. The standard deviation of the surface temperature at the end of the simulation are included as vertical grey bars. The data agrees well with \eqref{eqn:T0_temp} (solid black line).  }
        \label{fig:TempLimit}
\end{figure}

In all of the simulations presented here, $\Ra,\beta$ are sufficiently large that the system will become convectively unstable. This convection increases the vertical transport of heat. After an initial transient, both the average top temperature ($\Tzt$) and mean domain temperature ($\ThM$) decay nearly exponentially to their equilibrium values (see figure~\ref{fig:TempLimit}a). For the remainder of this paper, we will denote the stationary value of the surface temperature as $\ThO$, which was computed from the exponential fit (see the horizontal (black) dashed line in  figure~\ref{fig:TempLimit}a). 

After convection begins, the mean temperature profiles have a consistent structure; the centre of the domain is well mixed with sharp temperature gradients near the boundaries (see figure~\ref{fig:TempLimit}b). The feedback between the convective heat flux and the boundary condition results in a gradual, not instantaneous, decrease in $\Havg{\Theta}_{z=1}\to\ThO$. The upper and lower temperature gradients increase accordingly. Note that the upper temperature boundary layer is typically thinner than the bottom boundary layer due to the different boundary conditions, leading to a mean temperature that is less than the mean of the boundary values $\ThM < \Delta/2$. 

\subsection{ What is the heat transfer rate at the water surface? } \label{sec::Results_Nu}

As the surface temperature changes, so do the top and bottom temperature gradients, and, correspondingly, $\Nu$. At the onset of convection, $\Nu$ is higher than its asymptotic value ($\Nu_0=\lim_{t\to\infty}\Nu$). That is, the value of $\Nu$ decreases over time, with $\Nu_0$ collapsing onto the curve \eqref{eqn:NuScl}, with 
$C=0.138\pm 0.007,\ p=0.308\pm0.003$ 
(see figure~\ref{fig:TempLimit}c). For the remainder of this paper, we will select $C=0.14, p=0.31$ as the power-law coefficients for $\Nu$.

We note that the observed power-law is less than the optimal value for free-slip boundaries \citep[$p\approx \frac 13$, see][]{wen_steady_2020}, but greater than the simulated values with no-slip boundaries and fixed aspect ratio \citep[$p\approx 0.28$, see][]{waleffe_heat_2015}. The results of \citet{clarte_effects_2021} suggest that, in the absence of rotation, $p\approx 0.29$. The current simulations with a no-slip bottom boundary and a free-slip surface are within the range of previously reported values.


\subsection{What is the equilibrium surface water temperature?}

From the Nu scaling \eqref{eqn:NuScl} and the boundary condition \eqref{eqn:BC}, we can write down an implicit equation for the top boundary temperature in the large $\RaD$ limit,
\begin{gather}
    \frac{C\Ra^p}{\beta }  \sim \frac{\ThO}{\left( 1 - \ThO\right)^{1+p}}, \qquad \RaD\gg1. \label{eqn:T0_temp}
\end{gather}
Noting that, here, the mean heat flux $J \sim C\Ra^p\Delta^{1+p}$. In the limit, $J\to1$, we recover the diffusive solution \eqref{eqn:DiffSolution}.

We find good agreement between the measured values of $\ThO$ and \eqref{eqn:T0_temp} with the estimated $\Nu$ coefficients $C,p$ (See figure~\ref{fig:TempLimit}d). That is, the ratio $\beta/\Ra^p$ is the fundamental parameter that determines the equilibrium surface temperature. As discussed in the introduction, it is only reasonable to assume a fixed surface temperature when $\beta\gg\Ra^p$. 

For finite $\beta$, the surface temperature fluctuates. The standard deviation of surface temperature at the end of the simulations (averaged in space and time) are denoted by the grey errorbars in figure~\ref{fig:TempLimit}d and are provided in table~\ref{tab:CaseParameters}. For both large and small values of $C\Ra^p/\beta$, the surface temperature is constrained and the temperature variations are minimal. Larger variations of $\approx0.06$ are consistently found when $C\Ra^p/\beta = O(1)$, and are primarily produced by the emergent convective circulation cells (see figure~\ref{fig:diagram}). As the form of the surface temperature is different in three-dimensions \citep{busse_nonlinear_1980}, we do not pursue this further and will investigate the structure of the surface temperature distribution in future work. 

\subsection{How quickly does the system reach equilibrium?}

The convection induced by surface cooling results in a rapid decrease in the mean water temperature (figure~\ref{fig:TempLimit}a), which eventually reaches a steady state. How long does it take to reach steady state? Figure~\ref{fig:asymp_timescale}(a) is a plot of the temperature perturbation $\left( 1- \ThM\right)$ normalized by its asymptotic value as a function of time for all cases. We find a clear difference in the time to equilibrium between the different cases.

As mentioned previously, the internal water temperature is nearly uniform, with strong gradients at the boundaries (figure~\ref{fig:TempLimit}b). Modelling this temperature profile as piecewise linear, the mean heat equation \eqref{eqn:heatBudget} and boundary condition \eqref{eqn:BC} simplify to
\begin{gather}
\derivt{} \ThM  = \frac{1}{\sqrt{\Ra\ \Pr}} \left[\left(\frac{\Tzt - \ThM }{\delta_{top}}\right) - \left(\frac{\ThM - \Tzb}{\delta_{bottom}}\right)\right], \qquad \left(\frac{\Tzt - \ThM }{\delta_{top}}\right) = - \beta \Tzt.\label{eqn:linearTheta}
\end{gather}
Here, $\delta_{top},\delta_{bottom}$ are the top and bottom boundary layer thicknesses, and $\Tzb=~1$. We will attempt to simplify the equation for $\ThM$ in the large Rayleigh number limit. 

We begin by assuming that each boundary layer separately satisfies a Nusselt-Rayleigh number relationship. That is, the upper and lower boundary layers satisfy,
\begin{align}
    Nu_u &= \frac{1}{\ThM - \Tzt} \frac{\ThM - \Tzt}{\delta_{top}} \sim C_1 \left( Ra_0 \left(\ThM - \Tzt\right)\right)^p, \label{eq::Nu_u}\\
    Nu_l &= \frac{1}{\Tzb - \ThM} \frac{\Tzb - \ThM}{\delta_{bottom}} \sim C_2 \left( Ra_0 \left(\Tzb - \ThM\right)\right)^p, \qquad \Ra\to\infty. \label{eq::Nu_l}
\end{align}
As discussed in the introduction, we expect $C_1\ne C_2$ due to the different upper and lower boundary conditions. Rearranging, we can simplify equation \eqref{eqn:linearTheta} above to 
\begin{gather}
    \phi \derivt{} \ThM  = \left(1 - \ThM\right)^{1+p} - \epsilon \Theta_{z=1}, \qquad \left(\ThM - \Tzt\right)^{1+p} = \frac{\epsilon}{ \rfactor} \Tzt, \label{eq::linearMeanTemp}
\end{gather}
where we have defined the parameters 
\begin{gather}
    \phi = \frac{\sqrt{\Ra\ \Pr}}{C_2\Ra^p},\qquad \epsilon = \frac{\beta}{ C_2 \Ra^p},\qquad \rfactor = \frac{C_1}{C_2}. 
\end{gather}

We consider two extreme limits for $\epsilon$. As discussed above, in the $\epsilon=0$ limit, the surface boundary condition tends to an insulating boundary condition. Conversely, the $\epsilon\to\infty$ limit fixes the surface temperature, which is more similar to the classic Rayleigh-B\'{e}nard problem. Investigating these two limits, we will estimate the timescale to equilibrium that depends on $\phi$, $\epsilon$ and $\rfactor$.

\subsubsection{Small $\epsilon\ll1$}
As we increase $\Ra$ with fixed $\beta$, the parameter $\epsilon$ tends to 0. In fact, most of the simulations performed in this paper have $\epsilon<1$. However, in the limit $\epsilon=0$, the insulating boundary condition results in no temperature difference across the layer, and the Nusselt number relationships \eqref{eq::Nu_u}--\eqref{eq::Nu_l} are invalid. In this analysis, we assume that $\epsilon$ is small but sufficiently large that the system remains convective. 

We look for a perturbation series solution for $\ThM$ and $\Tzt$. Scaling through, we determine a solution in small $\epsilon$ as 
\begin{align}
    \ThM = 1 - \epsilon^\frac{1}{1+p} \left( x_{\epsilon,1} + \epsilon x_{\epsilon,2} + \dots\right),\qquad 
    \Tzt = 1 - \epsilon^\frac{1}{1+p} \left( y_{\epsilon,1} + \epsilon y_{\epsilon,2} + \dots\right), \qquad \epsilon\ll1.
\end{align}
For consistency, we further scale time as
\begin{gather}
    t =  \phi \epsilon ^\frac{-p}{1+p}\xi_\epsilon. \label{eq::tscl_ep0}
\end{gather}
At first order, \eqref{eq::linearMeanTemp} reduces for $x_{\epsilon,1}$ to, 
\begin{gather}
    \deriv{\xi_\epsilon}{ x_{\epsilon,1}} = 1 - x_{\epsilon,1}^{1+p}. \label{eq::meanT_ep0}
\end{gather}

We plot the numerical solution to \eqref{eq::meanT_ep0} in figure~\ref{fig:asymp_timescale}(b) along with a subset of the numerical simulations (indicated by the squares in figure~\ref{fig:asymp_timescale}(c)). Here, we select $C_2=0.24$ as an empirical parameter, without error bounds due to the low number of data points. Note that the choice of scale for $\epsilon$ determines both the amplitude and time scale for $x_{\epsilon,1}$. For the cases selected, the reduced model \eqref{eq::meanT_ep0} provides a reasonable approximation. Figure~\ref{fig:asymp_timescale}(c) plots the asymptotic value of $x_{\epsilon,1}$ for each of the numerical simulations. While each different $\Ra$ cases appear to plateau for $\epsilon\ll1$, the asymptotic values of $x_{\epsilon,1}$ are significantly lower for $\Ra=10^{6},10^{7}$, than for $\Ra=10^8,10^9$. We suggest that this change in amplitude may result from the corresponding $\Nu$ being closer to 1 in those cases and therefore \eqref{eq::Nu_u}--\eqref{eq::Nu_l} are perturbed. Nevertheless, \eqref{eq::tscl_ep0} provides a reasonable estimate for the equilibrium timescale for $\epsilon\ll1$ at large $\Ra$. 

\subsubsection{Large $\epsilon\to\infty$}

In the alternative limit, $\epsilon\to\infty$, the surface boundary condition is isothermal at $\Tzt=0$. We can then construct an asymptotic series for $\ThM$, $\Tzt$ as  

\begin{gather}
    \ThM = 1 - \left( x_{\infty,0} + \frac{1}{\epsilon} x_{\infty,1} + \dots \right) ,\qquad 
    \Tzt =  0 + \frac{1}{\epsilon} y_{\infty,1} + \dots , \qquad \epsilon\to \infty.
\end{gather}
Again, we use the timescale  
\begin{gather}
    t =  \frac{\phi}{\rfactor} \xi_\infty. \label{eq::tscl_epI}
\end{gather}
Scaling time by $\rfactor$ defines a timescale that is more consistent with \eqref{eq::tscl_ep0} as shown in figure~\ref{fig:asymp_timescale} (i.e. more similar to an e-folding time). To leading order, the heat budget \eqref{eq::linearMeanTemp} reduces to the simplified equation 
\begin{gather}
    \deriv{\xi_\infty}{ x_{\infty,0}} = - \frac{1}{\rfactor}\left(x_{\infty,0}\right)^{1+p} +  \left( 1 - x_{\infty,0} \right)^{1+p}. \label{eq::meanT_epI}
\end{gather}
In this large $\epsilon$ limit, the timescale and asymptotic value of $x_{\infty,0}$ are determined by $\rfactor$. We estimate $1/\rfactor\approx  0.6$.  The solution to \eqref{eq::meanT_epI} is plotted in figure~\ref{fig:asymp_timescale}(d) along with a subset of the numerical simulations (indicated by the squares in figure~\ref{fig:asymp_timescale}(e)). Figure~\ref{fig:asymp_timescale}(d) also includes the cases with $\Tzt$ fixed at 0. The equilibirum perturbation temperature $(1-\ThM_0$) decreases rapidly for $\epsilon<5$ as shown in figure~\ref{fig:asymp_timescale}(e). Plotting the $\epsilon>5$ cases, the solution to \eqref{eq::meanT_epI} agrees well with the numerical simulations in both amplitude and timescale.

\subsubsection{Matching fits}

\begin{figure}
    \centering
    \includegraphics[width=0.8\textwidth]{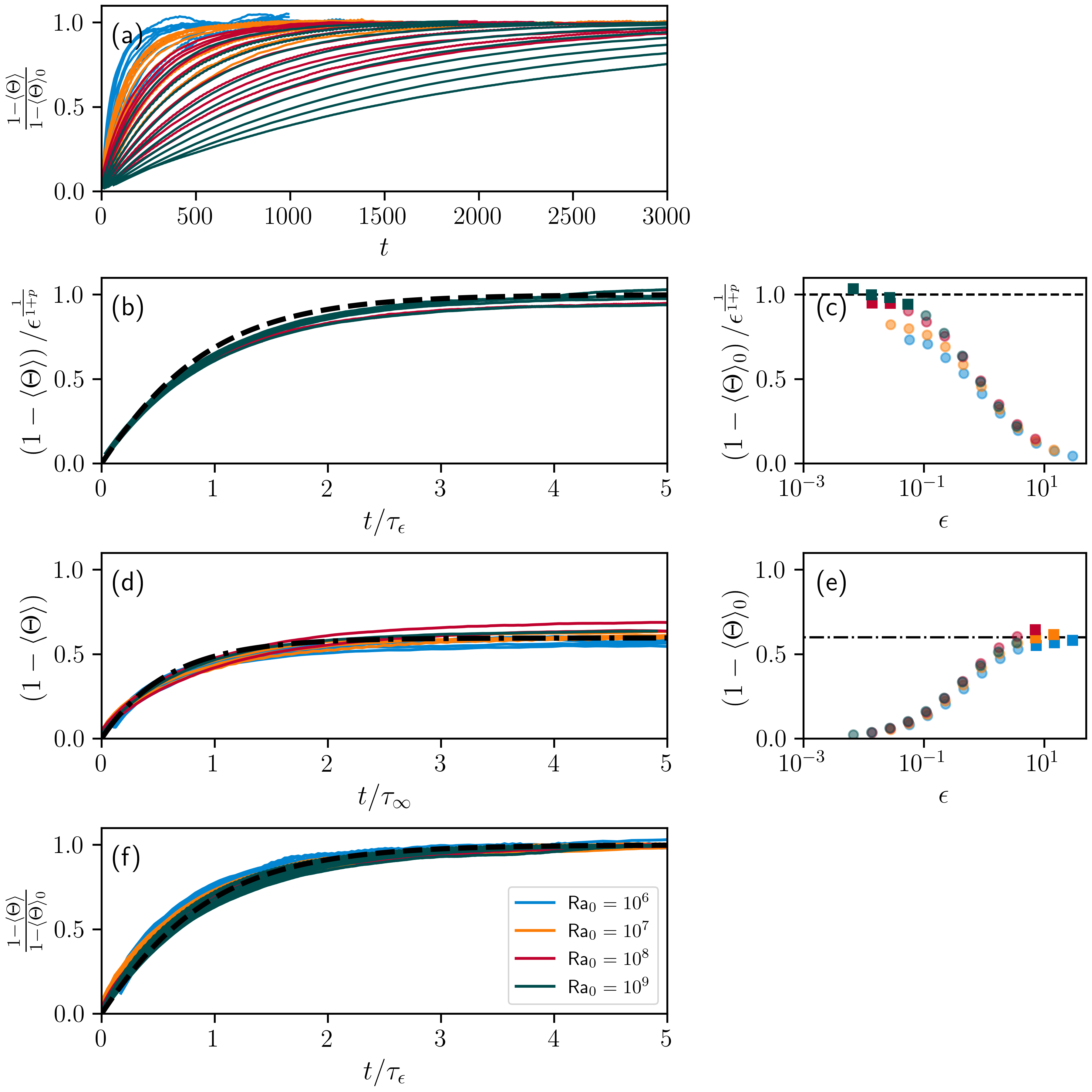}
    \caption{(a) Plot of the normalized temperature perturbation as a function of time. (b) Plot of the scaled temperature perturbation as a function of scaled time for $\epsilon\ll1$ with $C_2=0.24$. (c) Scaled asymptotic value of the temperature perturbations for all (non-fixed) cases as a function of $\epsilon$. Only cases denoted by a square are plotted in (b).  (d) Plot of the temperature perturbation as a function of scaled time for $\epsilon\to \infty$ with $1/\rfactor=0.6$. (e) Asymptotic value of the Temperature perturbation for all (non-fixed) cases as a function of $\epsilon$. The cases plotted with a square are plotted in (d) in addition to the cases with $\Tzt$ fixed at 0. (f) Plot of the normalized temperature perturbation for all cases at finite $\epsilon$ with scaled time. The solution to \eqref{eq::meanT_ep0} is included as a dashed lines in (b) and (f), while the solution to \eqref{eq::meanT_epI} is plotted as a dashed-dot line in (d). The timescales $\tau_\epsilon$ and $\tau_\infty$ are defined as in \eqref{eqn::tau}. }
    \label{fig:asymp_timescale}
\end{figure}

In developing the asymptotic solutions, we have defined two timescales ($\tau$) that were given by 
\begin{gather}
    \tau = \begin{cases}
    \tau_\epsilon = \phi \epsilon^{\frac{-p}{1+p}},& \epsilon\ll 1,\\
    \tau_\infty = \phi/\rfactor,                        & \epsilon\to \infty,\\
    \end{cases}
    \label{eqn::tau}
    \qquad \hbox{ where } \epsilon = \frac{\beta}{ C_2 \Ra^p} \quad \hbox{ and } \phi = \frac{\sqrt{\Ra\ \Pr}}{C_2\Ra^p}.
\end{gather}    
We suggest that we may be able to match these two solutions as $\tau = \phi \left(\epsilon^{\frac{-p}{1+p}} + f(\epsilon)\frac{1}{\rfactor}\right)$, where $f(\epsilon) \to 0$ for $\epsilon\ll0$, and $f(\epsilon)\to1$ as $\epsilon\to\infty$. Empirically, it appears that $\tau = \phi \epsilon^{\frac{-p}{1+p}}$ provides a reasonable estimate for the timescale in all simulated cases at finite $\epsilon$. Figure \ref{fig:asymp_timescale}(f) is a plot of the normalized temperature perturbation for all of the numerical simulations with finite $\epsilon$. Further, we find that the asymptotic solution where $\epsilon\ll1$ ( i.e. \eqref{eq::meanT_ep0}) provides a reasonable approximation for the shape of the mean temperature timeseries when normalized by their asymptotic values. Note that, contrary to the original derivation, this includes the cases where $\epsilon>1$ ( and finite). Based upon this surprising result, we suggest that $f(\epsilon)$ remains small even at moderate values of $\epsilon$.


\subsection{How vigorous is the convection?}


We now address our final question: How vigorous is the convection? We quantify this by looking at the kinetic energy of the system ($\TKE= \frac 12 \mathbf{u}\cdot \mathbf{u}$) in quasi-steady state. In that state, the $\TKE$ equation reduces to a balance of the buoyant production to dissipation. We have previously discussed the $\Nu$ scaling \eqref{eqn:NuScl}, which we will use to scale $\Vavg{w\Theta}$ as in \eqref{eqn:Nu}. Additionally, we can use the typical scaling for the viscous dissipation rate in terms of a turbulence length scale $l_\epsilon$. Writing these together, we have
\begin{gather}
    \Vavg{w\Theta} - \varepsilon = 0, \qquad \Vavg{w\Theta} \sim  \frac{C \Ra^p}{\sqrt{\Ra\ \Pr}} \Delta^{p+1}, \qquad \varepsilon \approx  \frac{\TKE^\frac 32 }{l_\epsilon},
\end{gather}
where $\varepsilon=2\sqrt{\frac{\Pr}{\Ra}} \Vavg{\underline{\mathbf{e}}: \underline{\mathbf{e}}}$ is the volume-averaged viscous dissipation rate based upon the rate of strain tensor $\underline{\mathbf{e}} = \frac 12 \left( \nabla \mathbf{u} + \left(\nabla \mathbf{u}\right)^T\right)$. 
%
We then solve for $\TKE$ as a function of $\Ra$ and $\Delta$, 

\begin{gather}
    \TKE = \left(l_{\epsilon} \frac{C \Ra^p}{\sqrt{\Ra\ \Pr}} \Delta^{p+1}\right)^\frac 23. \label{eqn:TKE}
\end{gather}
To close this equation, we need to determine $l_\epsilon$.

\begin{figure}
        \centering
        \includegraphics[width=\textwidth]{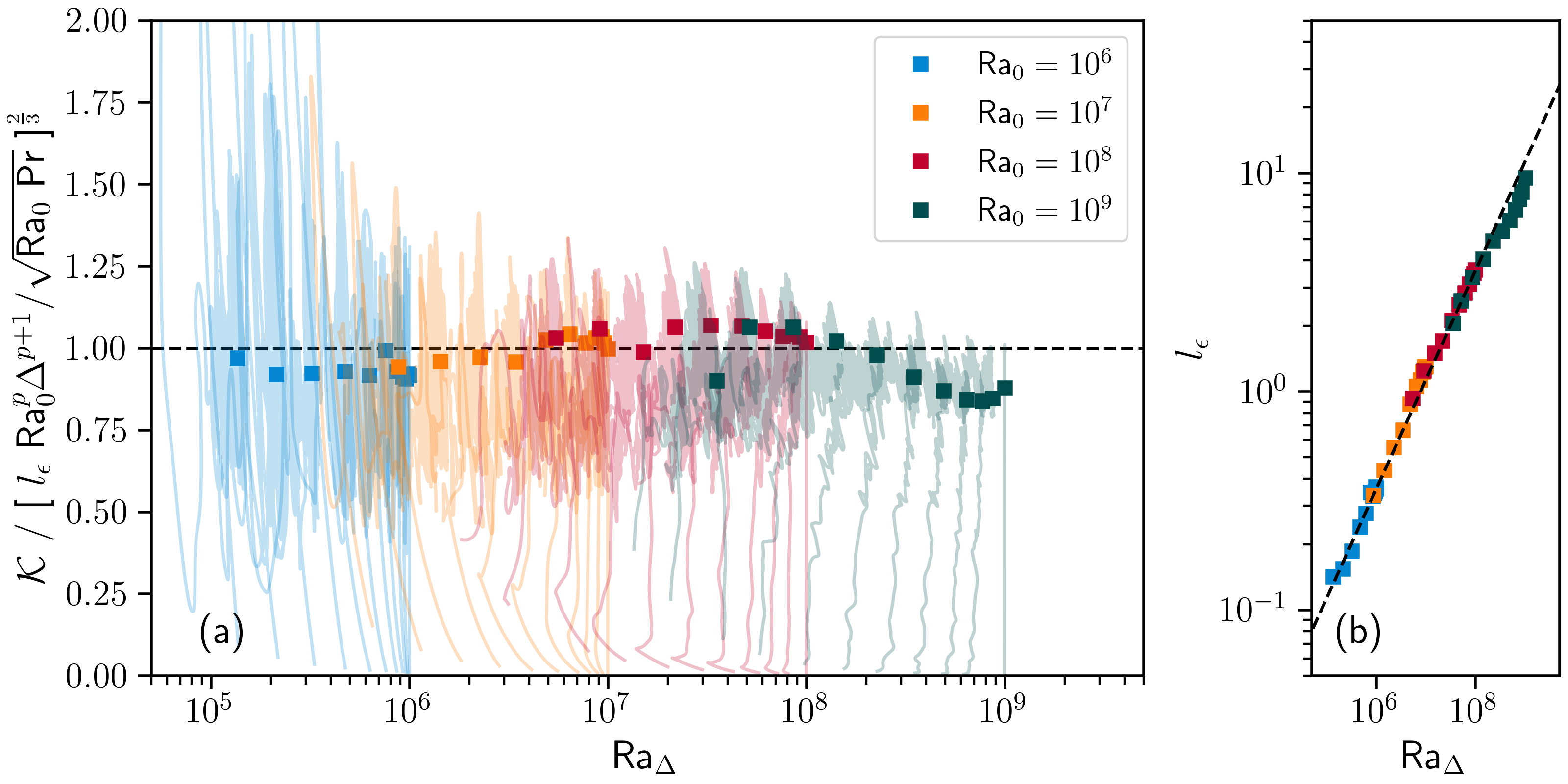}
        \caption{(a) The scaled kinetic energy as a function of $\RaD$. For this panel, $l_\epsilon$ is given by \eqref{eqn:le}. The lines denote the time evolution of $\TKE$ over the convective period of the simulation, while the squares denote the mean steady state value.  (b) Computed $l_\epsilon$ vs $\RaD$.}
        \label{fig:TKE}
    \end{figure}

Figure~\ref{fig:TKE}b is a plot of $l_\epsilon$, computed from the ratio of $\TKE$ and $\epsilon$, as a function of $\RaD$. We find that the data collapses well onto the curve,
\begin{gather}
    l_\epsilon = \left(5.3 \pm 0.6\right) \times 10^{-4\ }\RaD^{0.475 \pm 0.006}. \label{eqn:le}
\end{gather}
%
While there is some variation in the $\TKE$ around the predicted mean value, the scaling law \eqref{eqn:TKE} provides a reasonable estimate for the equilibrium $\TKE$ in this system (see figure~\ref{fig:TKE}a). Much of the observed variance about the horizontal line found in figure~\ref{fig:TKE}a could be removed by using a case specific value of $l_\epsilon$, rather than using the general parameterization \eqref{eqn:le}. However, the case-specific value does not eliminate oscillations about the mean. 

The relative strength of the fluid inertia to viscosity can be quantified through a Reynolds number Re. Using the predicted kinetic energy scaling, we can define the Reynolds number as \
\begin{gather} 
\hbox{Re} = \frac{\tilde U_{rms}H}{\nu} = \sqrt{\frac{2\Ra}{\Pr}} \left(l_{\epsilon} \frac{C \Ra^p}{\sqrt{\Ra\ \Pr}} \Delta^{p+1}\right)^\frac 13.
\end{gather}
In the present paper, we did not vary the Prandtl number and we do not know how $l_\epsilon$ changes with $\Pr$. Nevertheless, our kinetic energy scaling suggests that $\hbox{Re} \approx \Pr^q \Ra^{0.60} \Delta^{0.44}$ for some constant $q$ for large $\Ra$. Ignoring the Prandlt number dependence, the experimental work by \citet{lam_prandtl_2002} and \citet{niemela_turbulent_2006} suggest that $\hbox{Re}\sim \Ra^{0.4}$. The results from nonlinear numerical simulations by \citet{clarte_effects_2021} in a spherical shell with a similar boundary condition appear to scale $\hbox{Re}\sim \Ra^{0.5}$. However, the present results are more consistent with the optimal numerical predictions by \citet{wen_steady_2020} with $\hbox{Re}\sim \Ra^\frac{2}{3}$. We are unsure of the cause of this discrepancy, but one possibility is the two-dimensional nature of the present simulations. Future work will investigate the cause of this difference between studies.

\section{Conclusions}\label{sec:Conclusions}

Aquatic systems are coupled to their atmospheric forcing. We have extended the Rayleigh-B\'{e}nard problem to include this dynamic coupling. In particular, this coupling results in an extra parameter $\beta$, a scaled effective thermal conductivity that incorporates the effect of long-wave radiation, sensible heat loss, and evaporative heat loss. This model reduces to the typical Rayleigh-B\'{e}nard setup in the limit of $\beta\to\infty$. This extension to the dominant theory provides a relatively simple model to translate the results of Rayleigh-B\'{e}nard theory to environmental systems, such as lakes.

By defining an effective Rayleigh number, $\RaD = \Ra (1 - \Tzt)$, we have answered our four motivating questions:
\begin{enumerate}
    \item The surface heat transport is quantified through the Nusselt number $\Nu$, which we have shown to scale as
    \begin{gather}
        \Nu_0- 1 \sim C \RaD^{p}, \qquad \hbox{ where } C=0.138\pm 0.007,\ p=0.308\pm0.003.
    \end{gather}
    \item The equilibrium surface temperature ($\ThO$) is an implicit function of $\Ra$ and $\beta$, with 
    \begin{gather} \frac{C\Ra^p}{\beta }  \approx \frac{\ThO}{\left( 1 - \ThO\right)^{1+p}}. \end{gather}
    \item The system rapidly cools to its equilibrium value on an approximate timescale of 
    \begin{gather}
    \tau = \begin{cases}
    \phi \epsilon^{\frac{-p}{1+p}},& \epsilon\ll 1,\\
    \phi /\rfactor,                        & \epsilon\to \infty,\\
    \end{cases} \qquad \hbox{ where } \epsilon = \frac{\beta}{ C_2 \Ra^p} \hbox{ and } \phi = \frac{\sqrt{\Ra\ \Pr}}{C_2\Ra^p}.
    \end{gather}
    We estimate $C_2\approx0.24$ and $1/\rfactor\approx 0.6$.
    \item The kinetic energy of the induced convection scales as 
    \begin{gather}
    \TKE = \left(l_{\epsilon} \frac{C \Ra^p}{\sqrt{\Ra\ \Pr}} \Delta^{p+1}\right)^\frac 23, \qquad \hbox{where} \quad l_\epsilon = \left(5.3 \pm 0.6\right) \times 10^{-4}\ \RaD^{0.475 \pm 0.006}.
    \end{gather}
\end{enumerate}

While we have answered the questions put forth in this study, there remain many open questions that have not been addressed. In particular, the present simulations are two-dimensional. While outside the scope of this current study, future work will investigate how three-dimensional effects modify the present conclusions. Of particular interest would be the surface temperature variations due to the convection. While the surface temperature variations observed in this study were limited (see figure~\ref{fig:TempLimit}(d)), it would be interesting to study these variations in three-dimensional simulations that will likely have smaller scale structures at the water surface. In addition, the present model assumes that the atmospheric conditions were not varying. For forcing processes that change over a period much shorter than the equilibrium timescale $\tau$, there may be some interplay between the various timescales of interest. 

We also note that there appears to be weak curvature in Nu as a function of $\beta$. \citet{busse_nonlinear_1980} highlighted that, near the critical Rayleigh number, the structure of the flow is modified by the finite value of $\beta$. Later, \citet{ishiwatari_effects_1994} showed that two-dimensional cells have different preferred wavelengths in the two extreme cases of fixed temperature and fixed flux.  This has been further detailed by \citet{clarte_effects_2021} in nonlinear simulations of spherical shells. A detailed investigation of how the change in flow structure affects the curvature of Nu as a function of $\Ra$ and $\beta$ is an interesting avenue for future research.

We conclude by providing an estimate for the value of $\beta$ and $\Ra$ in real lakes. \citet{lepparanta_freezing_2015} provides an estimate for $\gamma \approx 20\ [\hbox{W m}^{-2} \hbox{K}^{-1}]$ for lakes in Southern Finland. If we assume that the entire lake is well mixed to a depth of 10 m with an atmospheric reference temperature of $\TO=283$ [K] and bottom water temperature of $\TzB=285$ [K] (e.g. in September), we estimate that $\beta \approx 350\ (\approx 2^{8.5})$ and $\Ra=O\left(10^{13}\right)$. These ballpark estimates are higher, but in a comparable range, to those considered here. While outside the scope of this present paper, future research would verify the current scaling laws in three-dimensions for these higher values of $\beta$ and $\Ra$. 



This manuscript discusses an extension to the usual Rayleigh-B\'{e}nard configuration that accounts for common surface cooling processes. 
Our hope is that this work will encourage a more direct comparison between Rayleigh-B\'{e}nard theory and field measurements in lakes and oceans.

    
    

\section*{Acknowledgement}

This work was funded by the Natural Sciences and Engineering Research Council of Canada (NSERC). 
I would like to thank Hugo Ulloa and Megan Davies Wykes for their helpful feedback on this paper, along with three anonymous referees.

\bibliographystyle{abbrvnat}
\bibliography{bib}

\end{document}

%% file: MacrosSymbols.tex
\newcommand{\diffz}{\frac{\partial}{\partial z} } 
\newcommand{\pderiv}[2]{\frac{\partial #2}{\partial #1} }
\newcommand{\deriv}[2]{\frac{\mathrm{d}#2}{\mathrm{d} #1} }
\newcommand{\derivt}[1]{\frac{\mathrm{d}#1}{\mathrm{d} t} }

\newcommand{\intV}[1]{\int_{V} #1 \  \mathrm{d}V}
\newcommand{\intA}[2][\partial V]{\int_{#1} #2 \  \mathrm{d}A^\prime}

\newcommand{\Vavg}[1]{\langle#1\rangle }
\newcommand{\Havg}[1]{\overline {#1}}

\newcommand{\Tzb}{\Theta_{z=0}}
\newcommand{\Tzt}{\Theta_{z=1}}
\newcommand{\mTh}{\overline \Theta}

\newcommand{\ThO}{\Theta_1}
\newcommand{\ThM}{\langle\Theta\rangle }

\newcommand{\TzH}{\tilde T_{\tilde z=\tilde H}}
\newcommand{\TzB}{\tilde T_{\tilde z= 0}}

\newcommand{\TO}{\tilde T_R}
\newcommand{\TA}{\tilde T_{Air}}

\newcommand{\Qs}{\mathbf{q}_{Surf}}

\newcommand{\TKE}{\mathcal{K}}
\newcommand{\Ra}{\hbox{Ra}_0}
\newcommand{\RaD}{\hbox{Ra}_\Delta}
\newcommand{\Nu}{\hbox{Nu}}
\newcommand{\rfactor}{\zeta}